\documentclass[a4paper,10pt,aps,prd,twocolumn,superscriptaddress,nofootinbib]{revtex4-1}

\usepackage{amsmath,amssymb,graphicx}
\usepackage{appendix}
\usepackage[T1]{fontenc}
\usepackage[utf8]{inputenc}
\usepackage{color}
\usepackage[unicode=true,pdfusetitle,bookmarks=true,bookmarksnumbered=false,bookmarksopen=false,breaklinks=true,pdfborder={0 0 0},backref=false,colorlinks=true]{hyperref}
\hypersetup{citecolor=blue,filecolor=blue,linkcolor=blue,urlcolor=blue}
\usepackage{pstool}

\newcommand{\cosmograph}{\textsc{CosmoGRaPH}}
\newcommand{\gevolution}{\textit{gevolution}}
\newcommand{\gramses}{\textsc{gramses}}

\begin{document}

\title{Numerical solutions to Einstein's equations in a {shearing}-dust Universe: \\ a code comparison}

\author{{Julian} Adamek}
\email[]{julian.adamek@qmul.ac.uk}

\affiliation{School of Physics and Astronomy, Queen Mary University of London, 327 Mile End Road, London E1 4NS, UK}

\author{{Cristian} Barrera-Hinojosa}
\email[]{cristian.g.barrera@durham.ac.uk}
\affiliation{Institute for Computational Cosmology, Department of Physics, Durham University, Durham DH1 3LE, UK}

\author{{Marco} Bruni}
\email[]{marco.bruni@port.ac.uk}

\affiliation{Institute of Cosmology and Gravitation, University of Portsmouth, Dennis Sciama Building, Portsmouth PO1 3FX, UK}
\affiliation{{INFN Sezione di Trieste, Via Valerio 2, 34127 Trieste, Italy}}

\author{{Baojiu} Li}
\email[]{baojiu.li@durham.ac.uk}
\affiliation{Institute for Computational Cosmology, Department of Physics, Durham University, Durham DH1 3LE, UK}

\author{{Hayley} J. Macpherson}
\email[]{h.macpherson@damtp.cam.ac.uk}

\affiliation{Department of Applied Mathematics and Theoretical Physics, Cambridge CB3 0WA, UK}
\affiliation{School of Physics and Astronomy, Monash University, Clayton, VIC 3800, Australia}

\author{{James} B. Mertens}
\email[]{mertens@yorku.ca}

\affiliation{Department of Physics and Astronomy, York University, Toronto, Ontario M3J 1P3, Canada}
\affiliation{Perimeter Institute for Theoretical Physics, Waterloo, Ontario N2L 2Y5, Canada}

\date{\today}

\begin{abstract}
A number of codes for general-relativistic simulations of cosmological structure formation have been developed in recent years. Here we demonstrate that a sample of these codes produce consistent results {beyond the Newtonian regime}. We simulate {solutions to Einstein's equations dominated by gravitomagnetism {---} a vector-type gravitational field that doesn't exist in Newtonian gravity} and produces frame-dragging, the leading-order post-Newtonian effect. We calculate the coordinate-invariant effect on intersecting null geodesics by performing ray tracing in each {independent code}. With this observable quantity, we assess and compare each code's ability to compute relativistic {effects}.
\end{abstract}

\maketitle

\section{Introduction}
The {flat} $\Lambda$ Cold Dark Matter ($\Lambda$CDM) model is the backbone of modern cosmology. {Originally proposed  in the context of the inflationary scenario \cite{Peebles:1984ge} and to accommodate for observations of structures on large scales \cite{Efstathiou:1990xe}, it has {emerged as} the concordance {cosmological model \cite{Spergel:2003cb,Tegmark:2003ud} after the discovery of the accelerating expansion of the Universe \cite{Riess:1998,Perlmutter:1999}}}.
{Theoretically, {$\Lambda$CDM} rests on three main pillars: {\it i)} 
 based on general relativity (GR) with a cosmological constant $\Lambda$,
 a Friedmann-Lema\^itre-Robertson-Walker (FLRW) {metric} is adopted as the description of the {Universe} on average, {on the assumption of large-scale statistical homogeneity and isotropy};
 {\it ii)} the relativistic perturbations of this background model are used to describe small inhomogeneities at large scales and early times, e.g.\ cosmic microwave background fluctuations; {\it iii)} Newtonian dynamics is used to model structure formation at late times and on small scales, where nonlinearity in the matter distribution is important.}

On these bases, $\Lambda$CDM successfully explains the majority of our cosmological observations in a surprisingly simple framework {\cite{Aghanim:2018eyx,Alam:2016hwk}. Yet $\Lambda$CDM faces a number of challenges, theoretical and observational. While a cosmological constant representing vacuum energy \cite{1931Natur.127..706L,Lemaitre1933} is the simplest possible form of dark energy, the measured %
{value is difficult to justify from a theory standpoint}
\cite{Adler1995,Weinberg1989,Martin2012}. With the continuous improvement  of cosmological observations, a number of tensions have started to emerge {\cite[see, e.g.][]{Buchert:2016}}, particularly between low and high redshift measurements of some cosmological parameters. {For example, a} {significant} {tension exists} between supernov\ae\ \cite{Riess2018} and cosmic microwave background measurements \cite{Aghanim:2018eyx} of the {present} {Hubble} expansion {rate}, $H_0$ \cite{Riess:2019cxk}. {T}he former depend{s} on calibration on the cosmic distance ladder \cite{Lemos:2018smw}, and the {latter depends} on assuming $\Lambda$CDM as cosmological model. Also assuming $\Lambda$CDM, a  tension is  present between high and low redshift observations of $\sigma_8$, the parameter measuring the growth of structures \cite{Battye:2014qga,Douspis:2018xlj}. Recently, some evidence for a spatially curved universe has been claimed \cite{Handley:2019tkm,DiValentino:2019qzk} and disputed \cite{Efstathiou:2020wem}, with some authors suggesting the possibility of a structure formation-induced curvature \cite{Bolejko:2017}. 
Motivated theoretically and because of these tensions, a number of alternatives to $\Lambda$CDM have been considered. These range from an interacting vacuum scenario \cite[see, e.g.][and references therein]{Salvatelli2014,Wang:2014xca,Martinelli:2019dau,Hogg:2020rdp}, to scalar fields \cite{Copeland2006} and modified  gravity models \cite[see e.g.][and references therein]{Clifton2012,Joyce2016, Ezquiaga2017,Frusciante:2019xia}{. However,} $\Lambda$CDM {is} still largely {preferred} when Bayesian model comparisons are carried out \cite{Heavens:2017hkr,Renk:2017rzu,Zhao:2017cud,Lazkoz:2018aqk,DiValentino:2019jae}.}

With the increasing precision of current and upcoming cosmological surveys {\cite{EuclidWeb,LSSTWeb,SKAWeb}, the  $\Lambda$CDM } model will be truly tested. {In view of these future observations and their target 1\% precision,} current state-of-the-art cosmological N-body simulations {of structure formation aim at the same precision in theoretical predictions. However, considering that these} N-body simulations are mostly based on the Newtonian approximation in lieu of full GR, {it is timely to address the possibility that some percent-level GR effects may be missed,} %
{potentially biasing the inferred likelihood of cosmological parameters.}
Understanding the role of general-relativistic effects on  observations will thus be crucial in correctly interpreting these precision data. 

While the extensions and alternatives to $\Lambda$CDM mentioned above explore new physics, some explore the inclusion of \emph{existing} physics that is neglected by the standard cosmological model. {Previous efforts to investigate the role of GR effects in numerical cosmology have included simplifying symmetries \citep[e.g.][]{Torres:2014,Rekier:2015}.}
In recent years, a number of general-relativistic codes  {\emph{with no assumed symmetries}} have been developed for cosmology, employing either a formally exact treatment of the metric
\cite{Loffler:2011ay,Giblin:2015vwq,Bentivegna:2015flc,Macpherson:2016ict,Clesse:2017,East:2017qmk,Daverio:2019gql} or an %
approximate scheme~\cite{Adamek:2015eda,Barrera-Hinojosa:2019mzo}. %
These tools provide new ways to study aspects of GR beyond the limited scope of
known analytic solutions and perturbative expansions around them. For instance, they have been applied to quantify gravitational back-reaction of small-scale structures \cite{Bentivegna:2012ei,Bentivegna:2013jta,Macpherson:2018akp,Adamek:2017mzb,Giblin:2018ndw,Macpherson:2019a}, light-cone projection effects \cite{Giblin:2016mjp,Bentivegna:2016fls,Giblin:2017ezj,Adamek:2018rru}, and the impact of relativistic species \cite{Adamek:2017grt,Adamek:2017uiq} on the evolution and observation of large-scale structure.
These codes have proven themselves reliable through comparisons to both linearized and exact GR solutions
(e.g.\ \cite{Adamek:2014qja,Adamek:2015hqa,Mertens:2015ttp,Bentivegna:2016stg}), and have in turn been used
to validate the applicability of traditional Newtonian simulations to cosmology in
a weak-field limit (e.g.\ \cite{Adamek:2016zes,East:2017qmk,East:2019chx}).

Here, we compare several codes within a controlled setup that features an artificially  large {gravitomagnetic vector potential as part of the metric, generating a}
frame-dragging effect. {Connected to rotation {of masses}, frame dragging {has been} measured {in the gravitational field of} the Earth \cite{gpb}.} {Gravitomagnetism and frame dragging are purely relativistic and absent from Newtonian gravity, a theory based on a single scalar potential (see however \cite{Bruni:2013mua} where this effect is {computed from a Newtonian code using} a {post-Friedmann approximation \cite{Milillo:2015cva}}).} 
There are only a few known analytic solutions
that exhibit frame-dragging {---} e.g. the Kerr and Kerr-Newman solutions  {---} although {the effect is ubiquitous in GR, {for example} in rotating neutron stars {\cite[see, e.g.][]{Berti:2004ny}}}. Numerical cosmological solutions
with large frame-dragging effects have also been studied \cite{Giblin:2019pql}.

In the limit where linear {cosmological} perturbation theory provides an accurate description of the behavior of a spacetime,  frame-dragging  is associated with the
presence of vector modes \cite{Bardeen:1980kt,Bruni:1992dg} and is well understood and under analytic control.
However, vector modes may also be sourced through nonlinear processes  {\cite[e.g.][]{Bruni:1996im,Matarrese:1997ay,Lu:2008ju,Thomas:2014aga,Jelic-Cizmek:2018gdp,Rampf:2016wom,Gressel:2019jxw}},
potentially interfering with our ability to measure phenomena such as primordial gravitational waves of sufficiently small amplitude \cite{Saga:2016cvt}. {They {can also produce  a gradient in the matter density field \cite{Ellis:1990gi}}.}
While such vector modes are not expected to be a dominant contribution to either the dynamics of our Universe or the propagation of information through it, simulating such a physical system nevertheless provides us with a way to explore the regime of validity of approximate
numerical and analytic models. At the nonlinear level, we must seek to validate both perturbation theory and
approximate numerical approaches against fully general-relativistic calculations.

In this work, we do precisely this. We describe and perform a comparison between linear theory, {simulations that use an approximate treatment of Einstein's equations to model nonlinear effects, and numerical relativity simulations that provide numerically exact solutions to Einstein's equations}. {{We} focus {our comparison} on the purely relativistic frame-dragging effect{. It} would {of course} be interesting to {study} other relativistic {effects} relevant for cosmology, e.g.\ in {nonlinear} structure formation or gravitational waves. {While these lie beyond the scope of this paper,} we emphasise that switching on gravitomagnetism goes a long way, because all relativistic degrees of freedom are excited once nonlinearity becomes relevant. On this basis, it seems reasonable to expect that the level of agreement among codes {in this test} should be a good indication of what to expect in other regimes.}

We present the initial data for our simulations in Sec{tion}~\ref{sec:initial_data}, describe the observables we compute 
in Sec{tion}~\ref{sec:observable}, {give an overview of the different computational frameworks in Section~\ref{sec:comp_frameworks}, discuss our results in Section~\ref{sec:discussion}} and conclude in Section~\ref{sec:conclusions}.

Unless otherwise stated, we use Latin indices to represent spatial indices, which take values $1, 2,$ and $3$, and Greek indices to represent space-time indices which take values $0, 1, 2$, and $3$, with repeated indices implying summation. We set the speed of light $c=1$.

\section{Initial data}
\label{sec:initial_data}
{GR admits a well-posed initial-value formulation {\cite[see, e.g.][]{Wald:1984rg}}. A consequence of this is that in cosmological simulations we don't need to specify an {\it a-priori} fixed background.}
We choose coordinates such that our initial Cauchy surface is {described by} a fixed coordinate time{, $t {= t_\ast}$. The {line element} in the {3+1} decomposition is} \cite{MAbook}
\begin{equation}
\label{eq:metric}
 g_{\mu\nu} dx^\mu dx^\nu = -\alpha^2 dt^2 + \gamma_{ij} \left(dx^i + \beta^i dt\right) \left(dx^j + \beta^j dt\right)\,,
\end{equation}
where $x^i \in \left\{x, y, z\right\}$ are coordinates on the three-dimensional space-like hypersurface, $\alpha$ is the lapse function, and $\beta^i$ is the shift vector. While we will set initial conditions non-perturbatively, we can draw a connection to linear perturbation theory both for intuition and a comparison. We choose the initial lapse, shift, spatial metric, and extrinsic curvature to be, respectively,
\begin{eqnarray}
\label{eq:ICalpha}
 \alpha_\ast &=& 1\,,\\ \label{eq:ICbeta}
 \beta^i_\ast &=& 0\,,
\end{eqnarray}
\begin{eqnarray} 
\label{eq:ICgamma}
 \gamma^\ast_{ij} &=& \begin{pmatrix}
                       1 & \frac{b}{H_\ast L} \cos\frac{2\pi y}{L} & 0 \\
                       \frac{b}{H_\ast L} \cos\frac{2\pi y}{L} & 1 + \frac{b^2}{H_\ast^2 L^2} \cos^2\frac{2\pi y}{L} & 0 \\
                       0 & 0 & 1
                      \end{pmatrix}\,,\\ \label{eq:ICKij}
 K^\ast_{ij} &=& -\begin{pmatrix}
                  H_\ast & \frac{b}{4 L} \cos\frac{2\pi y}{L} & 0 \\
                  \frac{b}{4 L} \cos\frac{2\pi y}{L} & H_\ast - \frac{b^2}{2 H_\ast L^2} \cos^2\frac{2\pi y}{L} & 0 \\
                  0 & 0 & H_\ast
                 \end{pmatrix}\,.~
\end{eqnarray}
This can be regarded as a snapshot at an initial time $t_\ast$ of an {\it exact perturbation} of a reference Einstein-de~Sitter {(EdS)} model {with  Hubble expansion  rate  $H_\ast$.} {Here,} $L$ is the characteristic length scale of the vector perturbation that also determines the size of the simulation volume (the initial conditions are compatible with periodic boundary conditions that identify $y=L$ with $y=0$), $b$ is a dimensionless amplitude, and the asterisk indicates that a quantity is evaluated on the initial Cauchy surface. {The surface has vanishing three-dimensional Riemann tensor ($\gamma_{ij}^\ast$ is the Euclidean metric in unusual coordinates) but non-trivial extrinsic curvature.} The connection with linear perturbation theory will become apparent shortly.

Having fixed the initial conditions for the metric we can proceed by solving the Hamiltonian and momentum constraint equations on the initial surface to obtain valid initial data for the matter. We assume that the matter can be described (at least initially) 
as a perfect fluid with vanishing pressure, such that the stress-energy tensor is given by
\begin{equation}
 T^{\mu\nu} = \rho_0 u^\mu u^\nu\,,
\end{equation}
where $\rho_0$ is the rest mass-energy density and $u^\mu$ is the four-velocity of the fluid. For collisionless matter the fluid description can break down at some point in the evolution due to stream crossing, and we will comment on this issue later.

The Hamiltonian and momentum constraint equations are, respectively,
\begin{eqnarray}
 {}^{(3)}\!R + K^2 - K_{ij} K^{ij} &=& 16 \pi G \alpha^2 \rho_0 \left(u^0\right)^2\,,\label{eq:hamiltonian}\\
 D_j \left(K^{ij} - \gamma^{ij} K\right) &=& 8 \pi G \alpha \rho_0 u^i u^0\,,\label{eq:momentum}
\end{eqnarray}
where the 3-curvature ${}^{(3)}\!R$ and covariant derivative $D_j$ are associated with the 3-metric $\gamma_{ij}${, and $K=\gamma^{ij}K_{ij}$}.
Together with the mass-shell condition $g_{\mu\nu} u^\mu u^\nu = -1$, this yields a closed system of equations from which we can determine $\rho_0^\ast$ and $u^\mu_\ast$. Solving this system, given the initial conditions \eqref{eq:ICalpha}, \eqref{eq:ICbeta}, \eqref{eq:ICgamma}, and \eqref{eq:ICKij}, we obtain
\begin{equation}
 \rho_0^\ast = 3 \frac{\left(16 H_\ast^2 L^2 - 3 b^2 \cos^2 \frac{2 \pi y}{L}\right)^2 - 64 \pi^2 b^2 \sin^2 \frac{2 \pi y}{L}}{128 \pi G L^2 \left(16 H_\ast^2 L^2 - 3 b^2 \cos^2 \frac{2 \pi y}{L}\right)}\,,
\end{equation}
\begin{equation}
 u_\ast^x = -\frac{8 \pi b \sin \frac{2 \pi y}{L}}{\sqrt{\left(16 H_\ast^2 L^2 - 3 b^2 \cos^2 \frac{2 \pi y}{L}\right)^2 - 64 \pi^2 b^2 \sin^2 \frac{2 \pi y}{L}}}\,,
\end{equation}
and $u_\ast^y = u_\ast^z = 0$.

It is worth noting at this point that values $b > 2 H_\ast^2 L^2 / \pi$ are unphysical, as they violate the weak energy condition {$\rho_0^\ast>0$ at $t_\ast$}. For $H_\ast L > \pi \sqrt{2/3}$, i.e.\ for {exact} perturbations outside the {Hubble} horizon, the physical range of $b$ is even more restricted, becoming bound{ed from above} by $b < 4 \sqrt{3 H_\ast^2 L^2 - \pi^2} / 3$.

All of the explicit expressions given so far are of course valid in the coordinate system we chose, {in particular with $\beta^i=0$}. We can therefore use these expressions directly to set the initial conditions in those simulations that use such coordinates. However, in some simulations we will instead use the coordinates of {the so-called} Poisson gauge in which vector perturbations are {completely} carried by the shift. In this gauge we denote the line element as
\begin{multline}
 \label{eq:PGmetric}
 \tilde{g}_{\mu\nu} d\tilde{x}^\mu d\tilde{x}^\nu = a^2(\tilde{t}) \bigl[-e^{2\psi} d\tilde{t}^2 + {\tilde{\gamma}^{ij} B_i B_j d\tilde{t}^2} - 2 B_i d\tilde{x}^i d\tilde{t}\bigr.\\ \bigl. + e^{-2\phi} \delta_{ij} d\tilde{x}^i d\tilde{x}^j + h_{ij} d\tilde{x}^i d\tilde{x}^j\bigr]\,,
\end{multline}
where {$a(\tilde{t})$ is the scale factor of the reference EdS model,} the shift $B_i$ is {the} transverse {gravitomagnetic vector potential}, $h_{ij}$ is transverse and traceless{, and $\tilde{\gamma}^{ij} = (e^{-2\phi}\delta_{ij} + h_{ij})^{-1}$ is the inverse of the spatial metric}.

For the particular initial data chosen, a closed expression for the coordinate transformation $x^\mu \mapsto \tilde{x}^\mu$ 
is given by
\begin{eqnarray}
\label{eq:ttrafo}
 \tilde{t} &=& t\,,\\
 \tilde{x} &=& x + \frac{b}{2 \pi} \left[\frac{1}{H_\ast} - \frac{3}{2} \left(t - t_\ast\right)\right] \sin\frac{2 \pi y}{L}\,,\\
 \tilde{y} &=& y\,,\\
 \label{eq:ztrafo}
 \tilde{z} &=& z\,,
\end{eqnarray}
for an infinitesimal $(t - t_\ast)$ around the initial Cauchy hypersurface. %
The metric variables in Poisson gauge are then initially given by
\begin{equation}\label{eq:ini_Poisson}
    B_i^\ast = -\begin{pmatrix}
                \frac{3 b}{4 \pi} \sin\frac{2\pi y}{L} \\
                0 \\
                0
        \end{pmatrix}\,,%
\end{equation}
and ${\psi_\ast =}~\phi_\ast = h_{ij}^\ast = 0$. %
{We note that {from the point of view of the Poisson gauge,} the extrinsic curvature %
{is only needed to provide the initial data for the propagating gravitational waves (i.e.\ the free part, or homogeneous solution, of $h_{ij}$). Due to the weak-field approximation, this part completely decouples from the remaining dynamics in \gevolution\ while it is neglected from the outset in \gramses\ --- which are the two codes detailed below that take their initial data in this coordinate system. Therefore the extrinsic curvature in Poisson gauge}
is not required for setting initial conditions {in this work}.}

{{From \eqref{eq:ini_Poisson} it is clear that} the dimensionless parameter $b$ measures the strength of the gravitomagnetic field on the initial hypersurface. {Neglecting} $b^2$ terms {in the above expressions} we obtain initial conditions for the first-order solutions in the two gauges. %
{We discuss this further in} Sec{tion}~\ref{subsec:linear_observable} below.}

\section{Behavior of observables}
\label{sec:observable}

We now want to construct an observable that can be used to ``measure'' the frame-dragging effect even in the nonperturbative case. {Consider} an observer comoving with the fluid and located at a point of symmetry where spacetime is invariant under a parity transformation. {Without} loss of generality we can {choose} the observer {to be} located at the origin $x_O = y_O = z_O = 0$. {Now} consider two events $A$ and $B$ on the initial hypersurface that emit a flash of light in all directions. Within the coordinate system that we introduced on the initial hypersurface, these events shall be located at $x_A = x_O - L, y_A = y_O, z_A = z_O$ and $x_B = x_O, y_B = y_O - L, z_B = z_O$. The null geodesics that connect each of these two events with the worldline of the observer get ``lensed'' by the frame-dragging effect {(see Figure \ref{fig:evolution} for an illustration)}. The ray coming from $A$ travels close to a plane of symmetry, while the ray coming from $B$ travels almost orthogonal to it. They will therefore be affected in different ways. One effect is that the angle $\vartheta$ between the two incoming rays is not exactly 90 degrees in the frame of the observer. The non-vanishing dot-product
\begin{equation}
\label{eq:dotproduct}
\cos\vartheta = \left.\frac{k^\mu_A e_\mu^i \delta_{ij} k^\nu_B e_\nu^j}{k^\mu_A u_\mu k^\nu_B u_\nu}\right|_O
\end{equation}
of the two direction vectors (in the observer's rest frame) is therefore an observable that directly relates to the frame-dragging effect. Here $u^\mu$ is the observer's four-velocity {(which coincides with that of the fluid in this case)}, $k^\mu_A$, $k^\nu_B$ denote the two null vectors of the incoming geodesics, and the $e^i_\mu$ are the basis vectors of the Fermi frame that, up to rotations, is fixed by the requirement that $u^\mu e_\mu^i = 0$ and $g^{\mu\nu} e_\mu^i e_\nu^j = \delta^{ij}$.

\subsection{Linear regime}
\label{subsec:linear_observable}

We first analyze a solution in the regime where the amplitude of the perturbation $b$ is small and a linear treatment {---} {i.e.\ a first-order expansion in $b$ about an {EdS} background} {---}  is therefore a good approximation. In this regime it is convenient to work with first-order gauge-invariant variables, an approach pioneered by Bardeen in his seminal work \cite{Bardeen:1980kt}. If matter has vanishing pressure, the first-order gauge-invariant vector mode decays like $1 / a^2$. Noting that $a \propto t^2$ if the Universe is matter dominated {(where $t$ is conformal time)}, we find that the linear solution {for a harmonic slicing and} with initial conditions given by eqs.~(\ref{eq:ICalpha})--(\ref{eq:ICKij}) is
\begin{equation}
 \alpha = \frac{t^2}{t_\ast^2}\,,\quad \gamma_{ij} = \begin{pmatrix}
                       \frac{t^4}{t_\ast^4} & \frac{b t}{2 L} \cos\frac{2\pi y}{L} & 0 \\
                       \frac{b t}{2 L} \cos\frac{2\pi y}{L} & \frac{t^4}{t_\ast^4} & 0 \\
                       0 & 0 & \frac{t^4}{t_\ast^4}
                      \end{pmatrix}\,.
\end{equation}
In Poisson gauge, the same linear solution reads
\begin{equation}\label{eq:ini_Poisson_lin}
 a = \frac{\tilde{t}^2}{\tilde{t}_\ast^2}\,,\quad B_i = -\begin{pmatrix}
                                                         \frac{3 b \tilde{t}^4_\ast}{4 \pi \tilde{t}^4} \sin\frac{2\pi y}{L} \\
                                                         0 \\
                                                         0
                                                        \end{pmatrix}\,,\quad \psi = \phi = h_{ij} = 0\,,
\end{equation}
thus in this gauge the entire metric first-order perturbation is encoded in the shift. In a generic gauge, Bardeen's gauge-invariant potential is a linear combination of the shift and the time derivative of the transverse-vector part of $\gamma_{ij}$. Through the momentum constraint this potential is sourced by one of the two matter vector velocity perturbations, namely the one representing the vorticity of $u^\mu$. Using the momentum constraint we can therefore relate the gauge-invariant    
 quantity in the two gauges, i.e.,
\begin{equation}
 \partial^j \partial_j B_i = \partial^j \partial_t \left(\gamma_{ij} / \alpha^2\right)\,.
\end{equation}
Here, the left-hand side represents the gauge-invariant vector mode of the metric in Poisson gauge and the right-hand side the same quantity in the other gauge.

We can also solve the null geodesic equations perturbatively. {As per \eqref{eq:dotproduct}, in order to obtain the direction vectors we contract} the null vectors at the observer with the basis vectors that provide local Fermi coordinates, which can be constructed perturbatively as well. {We note that to calculate \eqref{eq:dotproduct}, the photon four-vectors may also be directly projected into the observers local frame using the projection tensor $g_{\mu\nu} + u_\mu u_\nu$. However, we proceed using the basis vectors of the Fermi frame in this work.}
At leading order, the dot product becomes
\begin{equation}
\label{eq:dotproduct_linear}
    \cos\vartheta_\mathrm{lin} = b \left[\frac{H_\ast L + 6}{\left(H_\ast L + 2\right)^3} + \frac{32\pi^2}{H_\ast L} \int_0^{2 \pi} \!\!\!\frac{\cos \xi}{\left(4 \pi + H_\ast L \xi\right)^3} d\xi\right]\,.
\end{equation}

This first-order expression has several intuitive properties. First, it is directly proportional to the amplitude $b$ of the vector perturbation. Second, in the limit $H_\ast L \ll 1$ it asymptotes to $3 b / 4$ which is independent of $H_\ast L$. This makes sense because deep inside the horizon the time it takes for the light to reach the observer is much shorter than the dynamical time scale of the perturbation. The observable hence becomes insensitive to time evolution. Third, in the limit $H_\ast L \gg 1$ quite the opposite is true, and the asymptotic value is $2 b H_\ast^{-2} L^{-2}$. The signal gets damped because the vector mode decays significantly while the light travels through the spacetime.

\subsection{Nonlinear regime}
\label{subsec:nonlinear_observable}

Beyond linear order a minor complication arises because the two flashes of light do not arrive at exactly the same time. The angle between them remains uniquely defined in the observer's inertial reference frame, and thus
one needs to keep track of the rotation of that frame with respect to any coordinate system that is used for the calculation{. T}he arrival vector of the first ray must {therefore} be parallel transported along the observer's world line in order for the angles to be comparable. On the other hand, the time delay can be seen as another observable that is linked to the frame-dragging effect.

\begin{figure}
    \centering
    \includegraphics[width=\columnwidth]{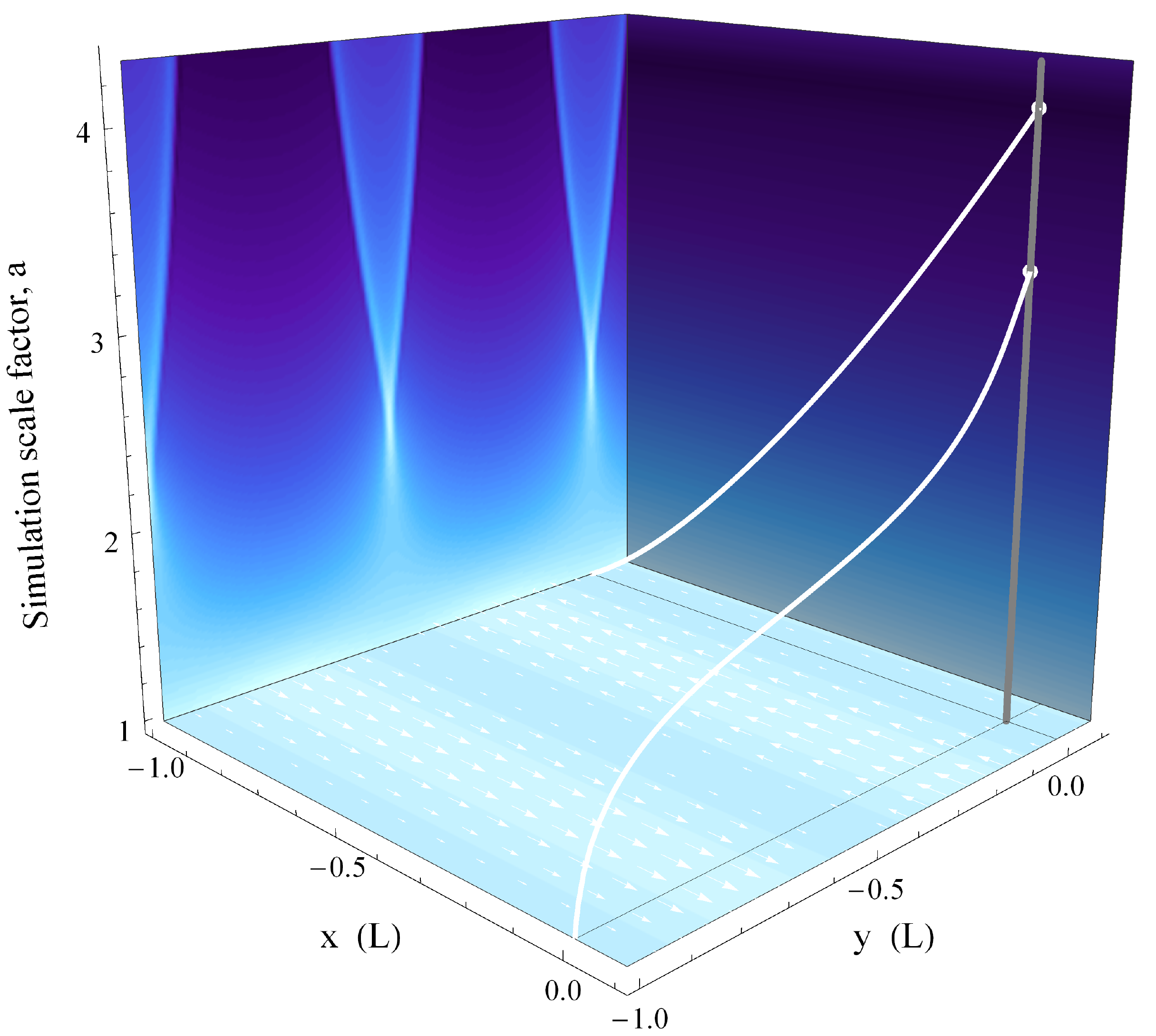}
    \caption{
    {White lines show} photon trajectories as they traverse the spacetime{, and} eventually intersect the observer's worldline, shown in grey. The colored panels in the background depict the ADM density as the simulation evolves after being initialized with a large amplitude and wavelength, $b=0.5$ and $H_* L = 2$. Lighter shades indicate higher densities, and darker lower densities. The arrows in the bottom panel depict the matter velocity on the initial surface.}
    \label{fig:evolution}
\end{figure}

In Figure~\ref{fig:evolution} we illustrate this visually, depicting photon trajectories as they traverse a spacetime with a large vector mode perturbation. Deflection of photons in this case can be manifestly seen, along with the time-delay. In the background, stream-crossings in the density field are observed as nonlinear collapse occurs. The ADM density in a gauge with a harmonic lapse condition and zero shift is plotted, which differs from the rest density at $\mathcal{O}(b^2)$, e.g.\ as noted in eq.~(\ref{eq:adm_dens}).

\section{Computational frameworks}\label{sec:comp_frameworks}

We compute {the} {observable} \eqref{eq:dotproduct} using four different relativistic codes: \gevolution, \gramses, the Einstein Toolkit (ET), and \cosmograph. Each code employs a different relativistic approach to evolving the initial conditions presented in Sec{tion}~\ref{sec:initial_data}{, as detailed below}. This study therefore provides a valuable comparison of these different computational methods in the context of a problem applicable exclusively to relativistic codes. 

We evolve the gauge-appropriate initial conditions (as per Sec{tion}~\ref{sec:initial_data}) with each code for several light-crossing times. This allows sufficient time for the light pulses to reach the observer even in the case of a large perturbation. For each code we perform 30 simulations with different initial conditions. Specifically, we choose ten values $H_\ast L = 2^n$ for $n\in\{-7,...,2\}$, and three initial perturbation amplitudes for each choice of $H_\ast L$, namely $b\in\{0.5,0.05,0.005\}\times H_\ast^2 L^2$. {This particular choice is motivated by the fact that $b / (H_\ast^2 L^2)$, much like the compactness parameter of compact objects, measures how extreme the matter configuration is. We remind the reader that on sub-horizon scales $b$ is bounded from above by $2 H_\ast^2 L^2 / \pi$, beyond which the matter configuration would no longer satisfy the weak energy condition. Some intuition can also be gained from considering the curl part of the momentum constraint in Poisson gauge. Here the Laplacian of the gravitomagnetic potential $B_i$ {---} whose initial amplitude is given by $b$ {---} is equal to the curl part of the momentum density ($\Delta^2 B_i = 16 \pi G a^2 \nabla\times\nabla\times T^0_i$ at leading order). Hence, a simple dimensional analysis shows that the latter has to approach the critical density as $b$ approaches the value $H_\ast^2 L^2$, up to some factors of order unity. The chosen} range of values {for $b$} allows us to sample {those} cases that we expect to be well within a linear regime, {as well as} those in which the {initial} perturbation has sufficient time to develop {into the nonlinear regime} during the time the light pulse takes to reach the observer.

{We trace} the path of the light pulses emitted at events $A$ and $B$ on the initial surface to the observer at the origin{, where we then} calculate the observable \eqref{eq:dotproduct}. To this end, {we integrate the geodesic equation in Wolfram Mathematica\footnote{\url{wolfram.com/mathematica}} (see Appendix~\ref{appx:photon_intgr}) for} the metric obtained {with each} code, until each light pulse reaches the observer{s position.} The corresponding boundary-value problem is solved using a shooting method (employing a built-in root-finding algorithm). In order to take into account the delay between the two observed signals, we parallel-transport one of the photon four-vectors along the observer's world-line before taking the dot product. Our ray-tracing method was validated by comparing three independent implementations, {including an extension to the \texttt{mescaline} code \cite{Macpherson:2019a}.}

{We confirm the numerical convergence of the observable \eqref{eq:dotproduct} for each code by simulating multiple spatial resolutions, see Appendix~\ref{appx:convergence} for details. We also compare the constraint violation in the numerical relativity codes \cosmograph\, and the ET in Appendix \ref{appx:constraint}.}

\subsection{\gevolution}
\label{sec:gevolution}

The public cosmological N-body code \gevolution\, is based on a weak-field expansion of Einstein's equations in Poisson gauge, which facilitates a vastly more efficient (yet ultimately approximate) computation in most cosmological settings \cite{Adamek:2015eda,Adamek:2016zes}. This is mainly due to the fact that non-relativistic particle motion allows for a superior convergence rate of the time integrator, making \gevolution\, an extremely useful tool in relativistic cosmology. {A crucial feature in this respect is the spin decomposition of the metric which separates the dynamical spin-2 field from the constraints. The latter evolve on time scales determined by the non-relativistic matter.} In our present setup, however, this advantage does not always play out, as we are exploring a parameter space that allows for highly relativistic particles. {Nonetheless, all the simulations presented here could be run in a few hours on a single desktop workstation.}

The vector mode of the metric, $B_i$, is kept only to linear order in \gevolution, but its source term, which is the spin-1 part of the momentum density, is treated non-perturbatively. To the extent that the solution maintains $\sqrt{\delta^{ij}B_iB_j} \ll 1$, which is true in cosmology and even for most of the parameter space studied here, this yields a self-consistent framework. Of course, the numerical solution will only be accurate up to corrections quadratic in $B_i$. We shall investigate the performance of this approximation in the regime of large $B_i$ in comparison with codes using numerical relativity.

{Our main results shown in the next section are based on simulations with periodic domains of $N=96$ grid points in each direction to sample the spacetime, and we also use the same number of particles. In order to study numerical convergence we also {performed} all simulations with $N=64$, and some with $N=48$.}

\subsection{\gramses}
\label{sec:gramses}

The recently introduced N-body code \gramses~\cite{Barrera-Hinojosa:2019mzo,Barrera-Hinojosa:2020arz} implements a constrained formulation of GR~\cite{Bonazzola-FCF:2004,CorderoCarrion:2008-1,CorderoCarrion:2011cq}, in which the Einstein equations are cast into a system composed of three hyperbolic equations for the evolution of tensor degrees of freedom, and a set of ten elliptic-type equations that explicitly include the constraints. Its current version neglects the hyperbolic part by using the conformal flatness approximation and solves the elliptic system using multigrid relaxation, which allows it to compute the two scalar and two vector modes of the metric. The code inherits the adaptive mesh refinements and massive parallelisation infrastructure from its parent code, {\sc ramses}~\cite{Teyssier:ramses}.

{In {\sc gramses}, the spatial coordinates are defined by the minimal distortion gauge (or generalized Dirac gauge) condition~\cite{CorderoCarrion:2011cq,Smarr1978:MDC}, $\partial_ih^{ij}=0$, where $h_{ij}$ corresponds to the deviation from a conformally flat spatial metric. Notice that this is generally different from the Poisson-gauge metric \eqref{eq:PGmetric}, in which $h_{ij}$ is both transverse and traceless{. Furthermore,} $\beta^i$ might carry both scalar (longitudinal) and vector degrees of freedom {whereas in Poisson gauge only the latter is allowed}. However, the initial data \eqref{eq:ini_Poisson} is actually fully compatible with the conformal flatness condition $h^\ast_{ij}=0$, so that the spatial coordinates at the initial hypersurface are equivalent in {these} two gauges, as well as the initial shifts. Moreover, in this code the time coordinate is fixed by a constant mean curvature (CMC) slicing condition, which \eqref{eq:ICKij} satisfies, and then \eqref{eq:ttrafo} also applies.} 

{{\sc gramses} obtains the {metric and extrinsic curvature components}
by solving elliptic-type equations on a mesh, {which means} the mesh resolution places a limit on the accuracy of its solutions through the discretisation error. In the results shown below we have used a mesh with $256^3$ cells, while we have found that using $128^3$ and $64^3$ cells leads to larger inaccuracies {even in the linear regime, where higher-order terms neglected by the conformal flatness approximation are subdominant}. {T}he same discretisation error {occurs} for all equations being solved, {and so} it can affect particle movements and thereby accumulate over time{. It is therefore} important to choose a sufficiently fine grid to suppress {this error}. Note that for finite differencing at a fixed order, the discretisation error is determined by the number of cells per side instead of the physical size of a cell.}

\subsection{Einstein Toolkit}\label{sec:ET}

The Einstein Toolkit\footnote{\url{einsteintoolkit.org}} is an open-source numerical relativity code built on the Cactus infrastructure \cite{Loffler:2011ay,Zilhao:2013}. Comprised of about 100 individual modules, the ET contains codes to evolve the vacuum Einstein equations using either the Baumgarte-Shapiro-Shibata-Nakamura (BSSN) \cite{Shibata:1995,Baumgarte:1999} or the conformal and covariant Z4 \cite{Alic:2012} formalism. In addition, it contains codes for relativistic (magneto-)hydrodynamics \cite{Baiotti:2005,Etienne:2015}, employing a fluid approximation for the matter distribution, and the \texttt{Carpet} adaptive mesh refinement and MPI driver \cite{Schnetter:2004}. Many in-built initial condition setups are also available, along with constraint violation and analysis modules, and simulation management.

The ET was first adopted for cosmological simulations in \cite{Bentivegna:2015flc,Macpherson:2016ict} (see also \cite{Bentivegna:2016stg}), and has since been shown to be a viable code for fully relativistic simulations of large-scale structure formation. Further studies have included structure formation growth rates \cite{Bentivegna:2015flc}, primordial gravitational waves \cite{Wang:2018a}, global backreaction and curvature \cite{Macpherson:2019a}, and the effect of small-scale inhomogeneities on the local expansion rate \cite{Macpherson:2018akp}. 

{While the ET is capable of evolving spacetimes in an arbitrary gauge, in this work the initial conditions are evolved using a harmonic lapse condition and zero shift. Because this is a fully covariant calculation, the final observable computed will be independent of the gauge used.} {Here we use {the BSSN formalism} to simulate three periodic, cubic domains with resolutions $N^3$, where $N=64,80$, and $96$, for each set of initial conditions. {Having multiple resolutions} allows us to quantify numerical errors for each simulation, details of which are given in Appendix~\ref{appx:convergence}.}

\subsection{\cosmograph}\label{sec:cosmograph}

Similar to ET, \cosmograph\, \cite{Mertens:2015ttp,cosmograph} is an open-source numerical relativity code employing the BSSN formulation of Einstein's equations in order to evolve the metric. It has incorporated SAMRAI \cite{Wissink:2001} in order to provide full adaptive mesh refinement and MPI capabilities. \cosmograph\, was developed to explore general relativistic effects in cosmological spacetimes, and to probe the applicability of novel numerical methods to cosmological problems. The framework can evolve matter sources including N-body systems, perfect fluids, and scalar fields; and can further perform raytracing through these spacetimes as they dynamically evolve in either a forward or time-reversed setting in order to compute various cosmological observables.

\cosmograph\, has been demonstrated capable of obtaining solutions with sufficient accuracy to robustly resolve relativistic corrections, down to the level of numerical precision. It has been used in a cosmological context to explore spacetimes in both weak-field \cite{Giblin:2015vwq,Giblin:2018ndw} and strong-gravity \cite{Giblin:2019pql} limits, and to explore observable properties of these spacetimes \cite{Giblin:2016mjp}. {Similar to ET, \cosmograph\ is capable of utilizing an arbitrary gauge through choice of lapse and shift, however for this work a harmonic slicing condition and zero shift are used.} {The periodic domain is simulated with resolution {$N_x=N_z=1$ in the x- and z-directions, and} $N_y=64$, $96$, and $128$ for each set of initial conditions} {with $N_p = N_y^2/8$ particles. The particle number is chosen to scale this way in order to obtain convergence of the physical field configurations and constraint violation \cite{Giblin:2018ndw}.}

\begin{figure}
    \centering
    \includegraphics[width=\columnwidth]{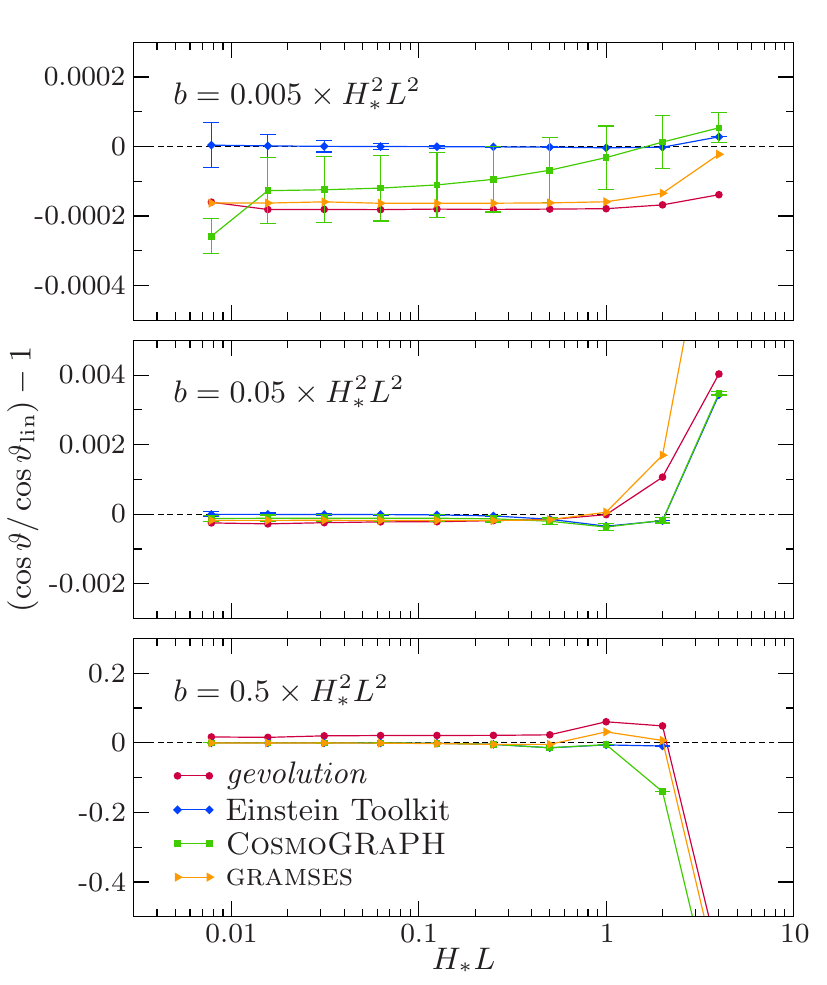}
    \caption{Numerical results obtained from \gevolution\ (red circles), the Einstein Toolkit (blue diamonds), \cosmograph\ (green squares) and \gramses\ (orange triangles). {Panels show} the relative difference between eq.~(\ref{eq:dotproduct}), evaluated numerically for simulations with different values of $b$ and $H_\ast L$, and the corresponding linear prediction given in eq.~(\ref{eq:dotproduct_linear}).}
    \label{fig:results}
\end{figure}

\section{Results \& Discussion}
\label{sec:discussion}

 Figure~\ref{fig:results} shows {results for} the observable \eqref{eq:dotproduct} {as computed from the different codes,} relative to the linear solution \eqref{eq:dotproduct_linear}, i.e.\ $({\rm cos}\vartheta/{\rm cos}\vartheta_{\rm lin})-1$, as a function of the dimensionless product $H_* L$; the length scale of the perturbation in units of the Hubble scale. Panels, top to bottom, show {three} different {sets of} initial perturbation amplitude{s}, $b \in \{0.005,0.05,0.5\}\times H_\ast^2 L^2$, respectively. Red circles show the results from \gevolution, blue diamonds show the results from the ET, green squares show the results from \cosmograph, and orange triangles show the results from \gramses, with dashed lines showing the linear solution for reference. Data and corresponding error bars for ET and \cosmograph\, were calculated using a Richardson Extrapolation, {i.e. the data points shown represent values extrapolated to $N\rightarrow \infty$}, see Appendix~\ref{appx:convergence} for details. 

In \gevolution, the weak-field approximation includes all terms in the expansion up to $\mathcal{O}(b)$, and so for large perturbations (and for $H_\ast L \gtrsim 1$) we do expect to see a difference with respect to the ET and \cosmograph. A similar statement can be made about \gramses, which, due to the conformal flatness approximation, neglects tensor modes that might be excited {during the evolution of the system} at $\mathcal{O}(b^2)$ and beyond.

The top panel of Figure~\ref{fig:results} shows results for the smallest-amplitude perturbation, i.e.\ $b=0.005\times H_*^2 L^2$. For this case we see agreement with the linear solution to within $0.03\%$ for all codes and for all values of $H_* L$.
{The ET provides the most numerically accurate solution in this regime where the perturbations remain linear. In this case, the fluid description implemented in the ET remains valid, which carries a smaller numerical error because it does not require smoothing over a particle distribution at each time step.}
To within numerical accuracy of the simulations performed, the only resolvable effect is a $0.003\%$ deviation from the linear solution for the ET with $H_* L = 4$ -- with the exception of one outlying \cosmograph\ point at the smallest $L$ and $b$ due to poor numerical convergence, discussed further in Appendix~\ref{appx:convergence}.

Especially at large values of $H_* L$ the performance of the linear prediction may seem surprising, but can be understood from the following considerations. First, working in Poisson gauge where the vector perturbation is in the shift, we can see that the linear expression for the shift is in fact exact on the initial hypersurface, {cf.~eqs.~\eqref{eq:ini_Poisson} and \eqref{eq:ini_Poisson_lin}}. The shift also appears only linearly in the geodesic equation (see Appendix \ref{appx:photon_intgr}), and the terms due to the lapse perturbation, which is formally $\mathcal{O}(b^2)$ {outside of the initial hypersurface}, would alone not lead to a deflection in the considered setup due to symmetry. Hence their effect appears only at one order higher, namely $\mathcal{O}(b^3)$. There are no new $\mathcal{O}(b^2)$ corrections as one moves away from the initial hypersurface because the matter dynamics are mainly due to inertia.

Considering the perturbations $b=0.05\times H_*^2 L^2$, shown in the middle panel of Figure~\ref{fig:results}, we see measurable deviation from the linear solution for values of $H_* L \gtrsim 1$. These are below $\sim 1\%$ in all codes, and the numerical relativity codes ET and \cosmograph\, agree within their quoted numerical accuracy. The results from \gevolution\, and \gramses\ show a qualitatively similar deviation from the linear prediction, and are well within expected truncation errors from higher-order terms of $\mathcal{O}(b^2)$ {(i.e.\ quadratic in the shift or higher, see {Section}\ \ref{sec:gevolution})} in the weak-field expansion. {However}{, in the case of \gramses\, the roughly constant deviation from the linear solution for $H_* L \lesssim 1$ is a result of the mesh discretization error (see {Section}~\ref{sec:gramses}).}  %

\begin{figure}
    \centering
    \includegraphics[width=\columnwidth]{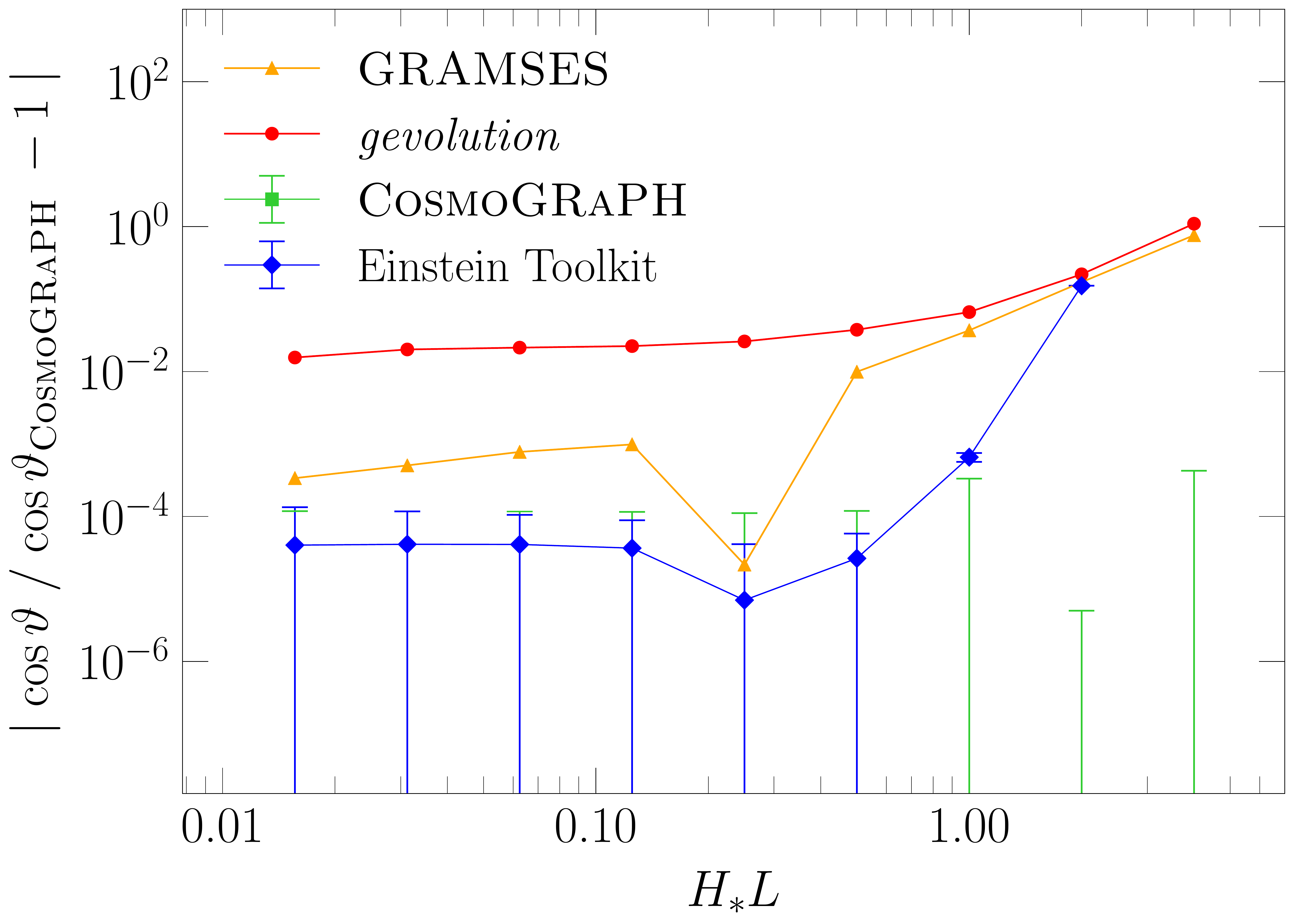}
    \caption{Relative difference between the observable \eqref{eq:dotproduct} measured in each code and that measured in \cosmograph\, for simulations with $b=0.5 \times H_*^2 L^2$. We use \cosmograph\, as a reference here as we expect it to produce the most reliable results in this extreme case. Red circles show residuals for \gevolution, orange triangles for \gramses, and blue diamonds for the Einstein Toolkit. Green lines show the error bars from the \cosmograph\, results.}
    \label{fig:results_rel_cosmograph}
\end{figure}

In the most extreme case {with} $b=0.5\times H_*^2 L^2$ {---} close to the limit set by the weak energy condition {---} shown in the bottom panel of Figure~\ref{fig:results}, we see the strongest deviations from the linear prediction. Incidentally, the linear prediction still holds to a good approximation for $H_* L \leq 1$, as confirmed by all four codes. In this regime, \textit{gevolution} shows a persistent $\sim 2\%$ error which is consistent with dropping terms of order $B_i \partial^2 B_i {\sim b^2 / L^2}$ from Einstein's equations{: compared to terms linear in $B_i$ their relative amplitude is indeed $\sim b/(H_\ast L)^2$ in some cases}.
The difference between \gevolution\, and the other three codes for %
all values of $H_* L$, is therefore still well within the expected $\mathcal{O}(b^2)$ truncation error from the weak-field expansion. 
{We have clipped the points for $H_* L = 4$ in the bottom panel of Figure~\ref{fig:results} to ensure deviations at smaller $H_* L$ can be resolved. In this case, we find deviation from the linear solution for \gevolution\, of $-0.619$, for \gramses\, of $-0.68$, and for \cosmograph\, of $-0.8183\pm0.0004$. {For the ET simulation with $H_* L = 4$, we} could not find {a null geodesic that connects either event A or B with the} observer, possibly indicating the presence of a horizon. We suggest this is a result of the fluid approximation used in the ET, since all other codes use a particle description and do not have this issue.

To highlight the difference between codes in the extreme regime shown in the bottom panel of Figure~\ref{fig:results}, we also plot the observable \eqref{eq:dotproduct} from each code relative to that calculated in \cosmograph\, in Figure~\ref{fig:results_rel_cosmograph}. Here we see \gevolution\, and \gramses\, agree to within $\sim 1\%$ and $ \lesssim 0.1\%$ for small values of $H_* L$, respectively. This difference grows to $\sim 1$ for the largest box size $H_* L = 4$. For $H_* L < 1$ the ET and \cosmograph\ agree within their numerical errors, however at $H_* L = 1$ we see a small deviation, which increases to $\sim 15\%$ for $H_* L = 2$.} In this case we expect stream crossing has occurred; a regime in which we no longer trust the fluid approximation implemented in the ET, and the $N$-body (Vlasov) description used in \cosmograph\, is more physically relevant. However, the fluid approximation is more numerically accurate than $N$-body per computational cost{, as can be seen clearly in the top panel of Figure~\ref{fig:results}}. This is due to the additional numerical error introduced by smoothing over the particle distribution at each time step in order to source the metric evolution, see e.g.\ \cite{Giblin:2018ndw}. Regardless, the inability of the fluid approximation to capture stream crossings implies that once these occur, the results from the ET should not necessarily be considered representative of collisionless matter.

\section{Conclusions}
\label{sec:conclusions}

In this work we have compared the computational approaches of four independent, relativistic, cosmological simulation codes. {Specifically, we {studied} a coordinate-invariant effect on null geodesics {that} is produced by a{n artificially} strong gravitomagnetic field.} In the process we have also explored the validity of the linear approximation for the frame-dragging effect. 

We summarize our main findings as follows:
\begin{itemize}
    
    \item For perturbations with amplitude $b=0.005\times H_*^2 L^2$, we find a match to linear theory within 0.03\% for all codes and for all box sizes studied here.

    \item For larger perturbations, with amplitudes $b \in \{0.05,0.5\}\times H_*^2 L^2$, we find agreement with linear theory within {0.1\%} and 1\%, respectively, for all {sub-horizon} box sizes {$H_\ast L \lesssim 1$. However, \gevolution\ has a persistent deviation of almost 2\% for the highest perturbation amplitude, which is nonetheless consistent with its approximation scheme.} %
    
    \item In \cosmograph, {the deviation} from linear theory {is} $\sim 80\%$ for $H_* L=4$ and $b=0.5\times H_*^2 L^2$. We expect \cosmograph\, to provide the most trustworthy results in this {extreme case; a regime in which $\gtrsim \mathcal{O}(b^2)$ effects are relevant \emph{and}} stream crossing occurs. 
    
    \item The weak-field approximation used in \gevolution\, agrees well with the numerical relativity codes for most cases studied here. Any deviations seen are well within the expected $\mathcal{O}(b^2)$ for all choices of parameters. {Due to the fact that Einstein's equations are second order, some of these corrections can scale as $\sim b^2/L^2$, which means that they survive even in the sub-horizon limit.} 
    
    \item {In \gramses, deviations from the linear solution in cases well into the linear regime are dominated by the mesh discretization error, while for larger perturbations %
    {deviations from other codes are mainly due} to the conformal flatness approximation, and are within the expected $\mathcal{O}(b^2)$ truncation error.}

    \item We find agreement within numerical uncertainties between the numerical relativity codes, ET and \cosmograph{, in most cases}. Exceptions are {those in} which stream crossing occurs, at which point the fluid description used in ET is no longer {resolving the full phase-space dynamics}, and \cosmograph\, provides a more {physically relevant} result.

\end{itemize}

The test performed here is unique in that it is applicable {\emph{only}} to codes which consider general-relativistic effects. Each code has differences in either its approximation of GR and/or numerical method. {This study therefore provides an important test of these approximations and their limits in describing nonlinear dynamics.} 

{A number of {other} relativistic effects would of course be interesting to investigate {using these codes. These include, but are not limited to, studying in detail the} collapse and virialisation of structures, the development and relevance of local spatial and Weyl curvature (electric vs.\ magnetic), coupling between small scales {---} where the matter distribution is very nonlinear {---} and the largest {sub-}horizon scales, gravitational waves, and the impact that any of these effects may have on {current and future cosmological} observations.

{The} codes {used here each} have {their} limitations, e.g.\ ET is {currently} limited to a fluid description of matter, and therefore {cannot} be used to study virialisation, while \gramses~uses the conformal flatness approximation and therefore {cannot} be used to study gravitational waves.
Comparing our codes in regimes where such effects may be relevant {therefore is} beyond the {scope} of {this} paper. Nonetheless, we {further} emphasise that by {inducing} a strong gravitomagnetic field, we have considered a regime where all relativistic degrees of freedom are excited {once} nonlinearity becomes relevant. 
{While} we leave the investigation of other relativistic effects to future studies, it seems reasonable to expect that these will not contradict the results we found here.}

\begin{acknowledgments}
We thank the anonymous referees whose comments contributed to the improvement of this paper. We thank the Sexten Center for Astrophysics\footnote{\url{http://www.sexten-cfa.eu}} for hosting a workshop on \textit{GR effects in cosmological large-scale structure} {in July 2018} that helped this project to get off the ground. JA acknowledges funding by STFC Consolidated Grant ST/P000592/1. 
MB is supported by UK STFC Grant No. ST/S000550/1.
HJM appreciates support received from the Herchel Smith Postdoctoral Fellowship Fund. The ET simulations in this work were performed on the MASSIVE HPC facility\footnote{\url{massive.org.au}} at Monash University, and the {SCIAMA} %
HPC cluster\footnote{\url{sciama.icg.port.ac.uk}} supported by the ICG, SEPNet and the University of Portsmouth. CB-H is supported by the Chilean National Commission for Scientific and Technological Research through grant CONICYT/Becas-Chile (No.~72180214). BL is supported by the European Research Council (ERC-StG-716532-PUNCA) and STFC Consolidated Grant (Nos. ST/I00162X/1, ST/P000541/1). The \gramses\ simulations used the DiRAC@Durham facility managed by the Institute for Computational Cosmology on behalf of the STFC DiRAC HPC Facility (\url{www.dirac.ac.uk}). The equipment was funded by BEIS via STFC capital grants ST/K00042X/1, ST/P002293/1, ST/R002371/1 and ST/S002502/1, Durham University and STFC operation grant ST/R000832/1. DiRAC is part of the UK National e-Infrastructure.
JBM would like to thank Tom Giblin and Glenn Starkman for conversations that helped shape this work. JBM also made use of the High Performance Computing Resource in the Core Facility for Advanced Research Computing at Case Western Reserve University. This research was supported in part by Perimeter Institute for Theoretical Physics. Research at Perimeter Institute is supported by the Government of Canada through the Department of Innovation, Science and Economic Development Canada and by the Province of Ontario through the Ministry of Research, Innovation and Science.
\end{acknowledgments}

\bibliography{grcodes}

\begin{appendices}

\section{Constraint violation in NR codes}\label{appx:constraint}

Here we analyse the constraint violation for the numerical relativity codes ET and \cosmograph\, for two select simulation cases. These codes are based upon hyperbolic formulations of Einstein's equations, which evolve the dynamical equations but do not explicitly enforce the constraint equations. The constraints \eqref{eq:hamiltonian} and \eqref{eq:momentum} are instead used as a diagnostic tool to check whether solutions have drifted too far from the physical constraint surfaces, i.e.\ to determine how well energy and momentum have been conserved in a general-relativistic sense.

Although this check is a good diagnostic, it does not {necessarily} imply validity of solutions, as numerical error can violate the dynamical evolution equations while still preserving constraints. For a timestep sufficiently small that the error in the dynamical evolution is small, remaining error will be dominated by truncation error when evolving vacuum or fluid solutions, or particle noise in the N-body case. In both cases, we can compute the rate at which the constraint violation in simulations converges to zero, and compare this to theoretical expectations.

In the case of stream-crossings, or caustics, it has been found that the constraint violation in the vicinity of a caustic will not converge in general \cite{Giblin:2018ndw,East:2019bwu}. This is due to the presence of a mild singularity in the vicinity of caustics, where curvature scalars can diverge, yet the metric remains in a weak-field limit and the spacetime is geodesically complete. In the N-body case, we therefore expect constraint violation to be well-behaved before stream-crossings, and poor after. We expect the constraint violation to be well-behaved at all times in the corresponding fluid limit (assuming all relevant scales are resolved, i.e. that all gradients in the fluid remain constant between resolutions).

\begin{figure}
    \centering
    \includegraphics[width=\columnwidth]{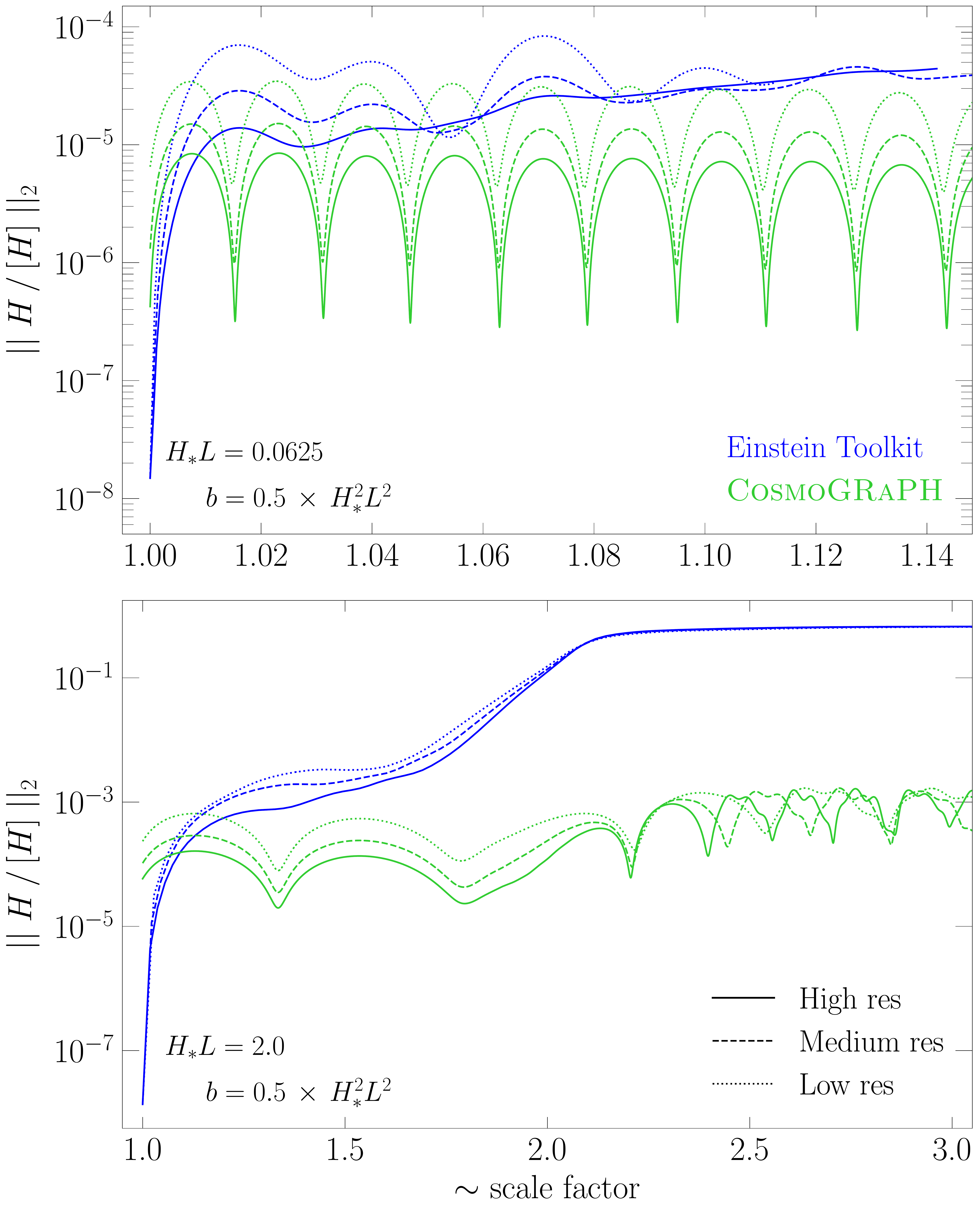}
    \caption{Hamiltonian constraint violation for the ET (blue) and \cosmograph\, (green), for $H_* L = 0.0625$ (top panel) and $H_* L  = 2$ (bottom panel). Both panels show simulations with amplitude $b= 0.5\times H_*^2 L^2$. Solid curves show the highest resolution, dashed curves the medium resolution, and dotted curves the lowest resolution. Here we show the L2 norm \eqref{eq:l2norm} of the constraint violation as a function of approximate simulation scale factor.}
    \label{fig:constraints}
\end{figure}

In Figure~\ref{fig:constraints} we show the constraint violation as a function of approximate scale factor of the simulation, for ET (blue curves) and \cosmograph\, (green curves). {We use the volume of the entire simulation domain, $V_\mathcal{D}$, to calculate the approximate scale factor, i.e.,
\begin{equation}
    a_\mathcal{D}(t) = \left(\frac{V_\mathcal{D}(t)}{V_\mathcal{D}(t_{\rm init})}\right)^{1/3},  
\end{equation}
where $V_\mathcal{D}\equiv\int_\mathcal{D} \sqrt{\gamma} \, d^3 X$, and $\gamma$ is the determinant of the spatial metric $\gamma_{ij}$.}

The top panel {of Figure~\ref{fig:constraints}} shows the simulation with $H_* L = 0.0625$, and the bottom panel shows $H_* L = 2$. Both simulations shown here have perturbation amplitude $b=0.5\times H_*^2 L^2$. Solid curves show the highest resolution, dashed curves the medium resolution, and dotted curves the lowest resolution. These sets of resolutions differ slightly between ET and \cosmograph, see Sections~\ref{sec:ET} and \ref{sec:cosmograph}, respectively, for details. Specifically, Figure~\ref{fig:constraints} shows the L2 error of the normalised Hamiltonian constraint violation, i.e.,
\begin{equation}\label{eq:l2norm}
    || H / [H] ||_2 = \frac{\sqrt{ \sum_i H_i^2 }}{\sqrt{ \sum_i [H]_i^2 }},
\end{equation}
where $H_i$ is the Hamiltonian constraint violation at grid cell $i$ (for an exact solution we have $H_i=0$ everywhere), and the normalisation is
\begin{equation}\label{eq:H_normalisation}
    [H] \equiv \sqrt{ \left({}^{3}R\right)^2 + \left(K^2\right)^2 + \left(K_{ij}K^{ij}\right)^2 + \left(16\pi G \rho \right)^2 },
\end{equation}
where $\rho = \alpha^2 \rho_0 (u^0)^2$ is the mass-energy density on the simulation hypersurfaces (not necessarily the rest-frame of the matter), and $[H]_i$ is \eqref{eq:H_normalisation} evaluated at grid cell $i$.

For the smaller box size, in the top panel of Figure~\ref{fig:constraints}, both simulations show convergence of the L2 error with an increase in resolution. For the larger box size $H_* L = 2$ in the bottom panel of Figure~\ref{fig:constraints}, the perturbation has more time to grow in a few light-crossing times of the box. For the N-body approach used in \cosmograph, convergence is found up until a stream-crossing occurs, at a scale factor of $a \sim 2$. After this time, inexact cancellation of numerically large (and formally infinite) contributions to the Hamiltonian constraint leads to a lack of convergence of constraint violation. For ET we see convergence until approximately the same time as \cosmograph\, in this case, after which the constraint violation is approximately the same at all resolutions, {and reaches order unity}. For this particular simulation, gradients are no longer consistent between resolutions and so we do not expect convergence of the constraints in general. After stream crossing, we do not expect the fluid approach used by ET to be representative of collisionless matter. {While convergence of the constraints diverges due to caustic formation,} convergence of the metric itself {and its first derivatives is still found, leading to convergent results for the observable as discussed in the next section.}

\section{Convergence and errors}\label{appx:convergence}

\subsection{Numerical convergence of observable}

{Here we check numerical convergence of the observable presented in Figure~\ref{fig:results} as a function of resolution for all codes. We must ensure that our numerical calculations provide results with a sufficient degree of precision that they are meaningful, i.e.\ that they are approaching a continuum limit solution.
We expect different rates of convergence for each code, due to different dominant error sources which depends on the numerical approximations made in each case.}

\begin{figure}
    \centering
    \includegraphics[width=\columnwidth]{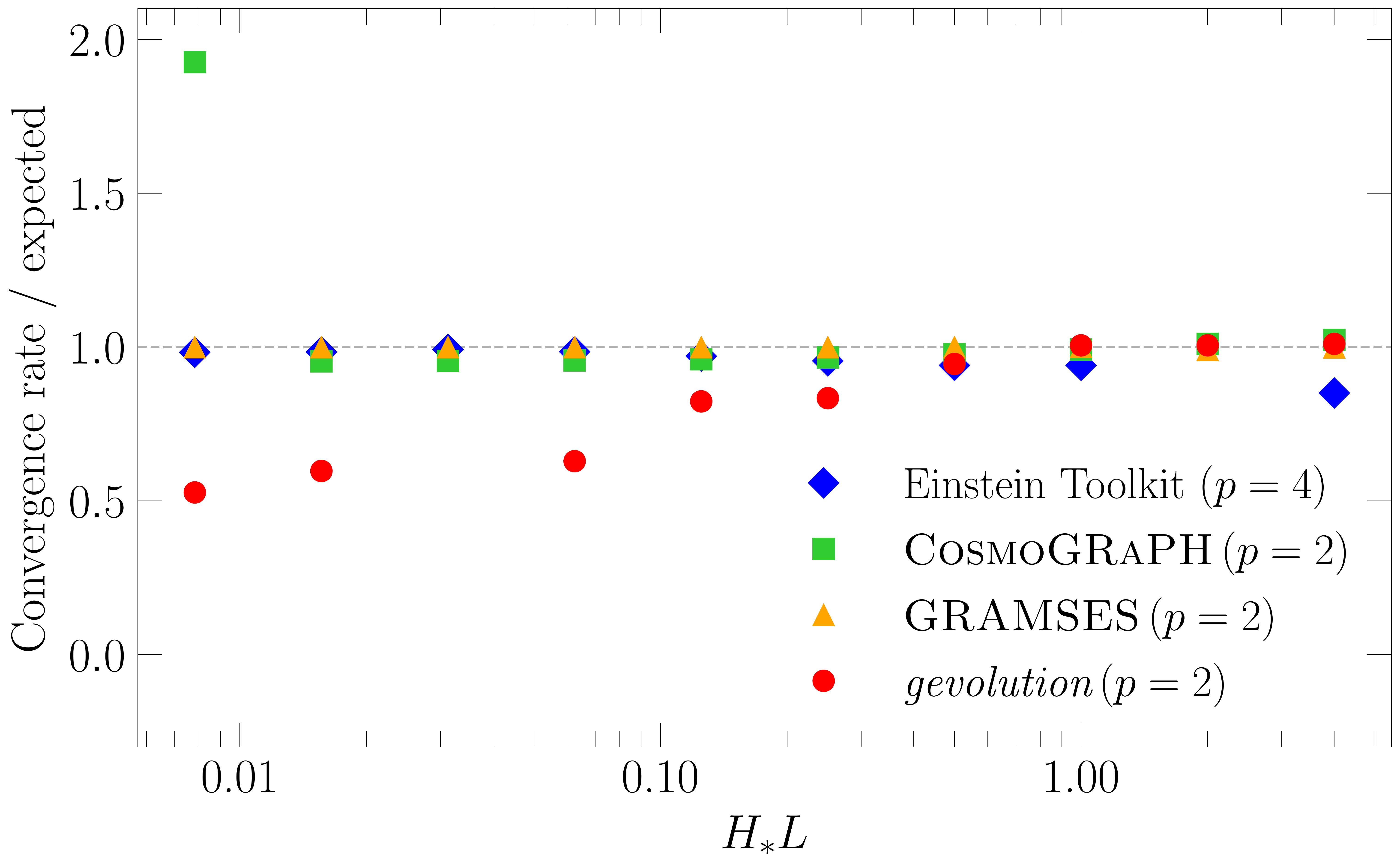}
    \caption{Convergence rate \eqref{eq:convergence} for the observable \eqref{eq:dotproduct} calculated in simulations with $b=0.005 \times H_*^2 L^2$. We show the convergence rate relative to its expected value~\eqref{eq:convergence-expected}, 
    for the Einstein Toolkit (blue diamonds), \cosmograph\ (green squares), \gramses\ (orange triangles), and \gevolution\ (red circles).{The order of the integration scheme, $p$, implemented in each code is indicated in the legend.}}
    \label{fig:convergence}
\end{figure}

{We compute} the convergence rate by evaluating the observable at {three} different resolutions, $\Delta x_1$, $\Delta x_2$, {and} $\Delta x_3$. For a method of order $p$, the error will be $\mathcal{O}(\Delta x^p)$, and the convergence rate of the observable angle $\vartheta$ is given by
\begin{equation}\label{eq:convergence}
    \mathcal{C} = \frac{\vartheta_{\Delta x_1} - \vartheta_{\Delta x_2}}{\vartheta_{\Delta x_2} - \vartheta_{\Delta x_3}},
\end{equation}
and the theoretical %
{convergence rate} is
\begin{equation}\label{eq:convergence-expected}
    \mathcal{C}_{\rm expected} = \frac{\Delta x_1^p - \Delta x_2^p}{\Delta x_2^p - \Delta x_3^p}\,.
\end{equation}
Figure~\ref{fig:convergence} shows the convergence rate relative to the theoretical convergence rate for each code, for the set of simulations in the top panel of Figure~\ref{fig:results}. {The expected convergence rate is calculated using \eqref{eq:convergence-expected} and the order of the integration scheme implemented in each code, $p$, as indicated in the legend}. For this small perturbation, we expect all results to match the linear solution to a good approximation, and so it is an ideal case in which to test numerical convergence {(although we have confirmed convergence for larger $b$ as well)}. {From Figure~\ref{fig:convergence} we see all codes give close to their expected numerical convergence rate for all values of $H_* L$. {In \gevolution\ we see a drop to first-order convergence inside the horizon. This is possibly due to the fact that we are approaching a quasistatic limit in which the error in the time integrator becomes subdominant. The elliptic constraint for the gravitomagnetic potential is solved by inverting a first-order finite-difference Laplacian, which may become the dominant source of error in this regime.}
We note a few peculiar convergence values {in Figure~\ref{fig:convergence}} for \cosmograph, \gevolution, and ET, at $H_* L = 0.0078125, 0.03125,$ and $2$, respectively. {In the case of \cosmograph,\ we believe this is due to truncation error surpassing error introduced from particle noise, leading to a different convergence rate than expected, and thus effective method order $p$; this additionally results in a poorly extrapolated data point and error bar as seen in Figure~\ref{fig:results}.}
{The remaining} simulations appear to show normal numerical convergence for, e.g., the constraint violation in the case of ET, and so we believe the simulations themselves are providing reliable results.
We therefore suggest that the non-convergence of the observable is related to the root-finding algorithm implemented in the ray tracing code used, but we do not investigate this further since all other points show good convergence.} 

\subsection{Error calculation}

{For ET and \cosmograph, simulations were run at three different spatial resolutions for each value of $b$ and $H_* L$. This allows us to quantify the numerical error on the observables we compute from these simulations{, as well as to estimate values in the $N\rightarrow\infty$} limit using a Richardson extrapolation. The values and error bars are calculated by fitting curves consistent with the expected order of convergence of the scheme used, and extrapolating to find the continuum-limit solution. The numerical error bars shown in Figure~\ref{fig:results} are computed using differences between numerical values and the extrapolated, continuum-limit solutions in the case of ET{, and using the distribution of extrapolated values in the case of \cosmograph.}
}

\section{Initial conditions with BSSN variables}

The BSSN conformal factor and extrinsic curvature trace are given by \cite{Shibata:1995,Baumgarte:1999}
\begin{eqnarray}
    \phi_\ast^{\rm BSSN} & = & 0 \\
    K_\ast & = & -3H_\ast\,.
\end{eqnarray}
Because {the determinant of the spatial metric} $\gamma^{*} = 1$, the conformal BSSN metric is given by $\bar{\gamma}_{ij}^{\ast} = \gamma_{ij}^*$. The conformally related trace-free part of the extrinsic curvature is given by
\begin{equation}
    \bar{A}_{ij}^{*}=3\begin{pmatrix}0 & \frac{3b}{4L}\cos\frac{2\pi y}{L} & 0\\
\frac{3b}{4L}\cos\frac{2\pi y}{L} & \frac{3b^{2}}{2H_{\ast}L^{2}}\cos^{2}\frac{2\pi y}{L} & 0\\
0 & 0 & 0
\end{pmatrix}\,,
\end{equation}
and the conformal, contracted Christoffel symbol is

\begin{equation}
    \bar{\Gamma}^{i}_{\ast}= \left(\begin{array}{c}
-\frac{2\pi b}{H_{\ast}L^{2}}\sin\frac{2\pi y}{L}\\
0\\
0
\end{array}\right)\,.
\end{equation}
Lastly, the 3+1/ADM density is given by
\begin{equation}
\rho^{*}_{\rm ADM} = \frac{3 H_*^2}{8 \pi {G} }-\frac{9 b^2 \cos ^2\left(\frac{2 \pi  y}{L}\right)}{128 \pi {G} L^2}\,,
\label{eq:adm_dens}
\end{equation}
and relativistic gamma factor $W_\ast=\alpha_\ast u_\ast^0$
\begin{equation}
W_\ast = \frac{16 H_*^2 L^2-3 b^2 \cos ^2\left(\frac{2 \pi  y}{L}\right)}{\sqrt{\left(16 H_*^2 L^2-3 b^2 \cos ^2\left(\frac{2 \pi  y}{L}\right)\right)^2-64 \pi ^2 b^2 \sin ^2\left(\frac{2 \pi  y}{L}\right)}}
\end{equation}

\section{Photon integration}\label{appx:photon_intgr}

The geodesic equations parallel transport velocity vectors in the direction of the velocity,
\begin{equation}
k^\mu \nabla_\mu k^\nu = 0\,,
\end{equation}
for a photon 4-vector $k^\mu$, or for ordinary matter with $k^\mu \rightarrow u^\mu$. In order to integrate this numerically, we can cast this expression into a 3+1 form conducive to numerical integration, and for which the effects of frame-dragging due to a nonzero shift are transparent,
\begin{eqnarray}
\label{eq:3+1geodesic}
\frac{dk_{i}}{dt} & = &-\alpha k^{0}\partial_{i}\alpha+k_{j}\partial_{i}\beta^{j} - \frac{1}{2k^0}k_{j}k_{k}\partial_{i}\gamma^{jk}\nonumber \\
\frac{dx^{i}}{dt} & = &\gamma^{ij}\frac{k_{j}}{k^{0}}-\beta^{i}\,.
\end{eqnarray}
Lastly, the observable, eq.~(\ref{eq:dotproduct}), can be rewritten without a reference to basis vectors for an observer at rest, $u^\mu \propto (1,\vec{0})$, as
\begin{equation}
\label{eq:covariant_observable}
    \cos\vartheta =\left. \frac{\gamma^{ij}k_{i}^A k_{j}^B}{\sqrt{\gamma^{ij}k_{i}^A k_{j}^A}\sqrt{\gamma^{ij}k_{i}^B k_{j}^B}}\right\vert_O\,,
\end{equation}
provided the vectors are observed simultaneously. In order to obtain this simple expression we also used the fact that $\beta^i$ vanishes at our observer location due to symmetry. For non-simultaneous arrivals, one vector will need to be parallel transported along the observer's trajectory.
{T}he parallel transport equation can be written
\begin{eqnarray}
    u^\mu \nabla_\mu k^{\nu} & = & 0 \nonumber \\
\rightarrow \frac{dk_{i}}{dt}&= & -\alpha k^{0}\partial_{i}\alpha+k_{j}\partial_{i}\beta^{j}-\frac{1}{2u^{0}}u_{l}k_{j}\partial_{i}\gamma^{lj} \nonumber \\
& & +\alpha\left(\frac{u_{k}}{u^{0}}k^{0}-k_{k}\right)K_{i}^{k}\,,
\end{eqnarray}
where the last term here is new compared to eq.~(\ref{eq:3+1geodesic}), and the second-last term now is sensitive to the observer's velocity.
In terms of $3+1$ variables, and for the metric and observer considered in this work, and for a gauge choice that respects symmetry of the problem (so $\partial_i \alpha = 0$ at $y=0$), this expression simplifies to
\begin{equation}
    \frac{dk_i}{dt} = k_j \partial_i \beta^j -\alpha k_j K^j_i\,,
\end{equation}
which is integrated purely in time at $y=0$. The observable, eq.~(\ref{eq:covariant_observable}), is also evaluated using the metric at the time the second ray arrives.

\end{appendices}

\end{document}